\documentclass{emulateapj}
\usepackage{graphicx}

\shorttitle{Unified Radio/Optical Catalog}
\shortauthors{Kimball \& Ivezi\'{c}}

\begin{document}
\bibliographystyle{apj}

\title{An Updated Version of the Unified Radio Catalog: A Multi-Wavelength Radio and Optical Catalog of Quasars and Radio Galaxies}
\author{Amy E. Kimball\altaffilmark{1}}
\author{\v{Z}eljko Ivezi\'{c}\altaffilmark{2}}

\altaffiltext{1}{{\tt akimball@nrao.edu}  National Radio Astronomy Observatory,
  Socorro, NM, 87801, USA}
\altaffiltext{2}{Department of Astronomy, University of Washington, Box 351580, Seattle, WA 98195-1580}

\begin{abstract}
We present a catalog of millions of radio sources, created by consolidating large-area radio and optical surveys GB6 (6cm), FIRST (20cm), NVSS (20cm), WENSS (92cm), VLSSr (4m), and SDSS DR9 (optical). The region where all surveys overlap covers 3269 deg$^2$ in the North Galactic Cap, and contains $>$160,000 20-cm sources, with about 12,000 detected in all five radio surveys and over one-third detected optically. Combining parameters from the sky surveys allows easy and efficient classification by radio and optical morphology and radio spectral index. The catalog is available at {\tt http://www.aoc.nrao.edu/\textasciitilde{}akimball/radiocat.shtml}.
\end{abstract}

\section{An updated catalog}

We present an updated version of the unified radio catalog published by \citet[][hereafter KI08]{ki08}. That catalog comprised sources detected at 20~cm by the Faint Images of the Radio Sky at Twenty cm survey \citep[FIRST;][]{first} and/or the NRAO-VLA Sky Survey \citep[NVSS;][]{nvss}, with supplemental data (when available) from the Green Bank 6-cm survey \citep[GB6;][]{gb6}, at 92~cm from the Westerbork Northern Sky Survey \citep[WENSS;][]{wensscat}, and in the optical by the Sloan Digital Sky Survey (SDSS) Sixth Data Release \citep[DR6][]{dr6}. 

The main addition to the catalog is the inclusion of 4-m (74-GHz) data detections (when available) from the VLA Low-Frequency Sky Survey Redux\footnote{http://www.cv.nrao.edu/vlss/VLSSlist.shtml} \citep[VLSSr;][]{vlssr}. Additionally, the latest versions of FIRST (updated 24 Feb 2012), and the final version of NVSS (version 41) are included.  The latest data reductions from those surveys have resulted in modified source lists (especially at low signal-to-noise) and small variations in measured parameters; however, the overall properties of the radio source populations in the unified catalog have not changed significantly. Finally, for the updated catalog we have incorporated data from the Ninth Data Release \citep[DR9][]{dr9} of the SDSS instead of the earlier DR6. As a result, the number of sources in the radio catalog that have optical spectra has increased by almost 60\%.

\begin{figure*}[t]
\includegraphics[width=7in]{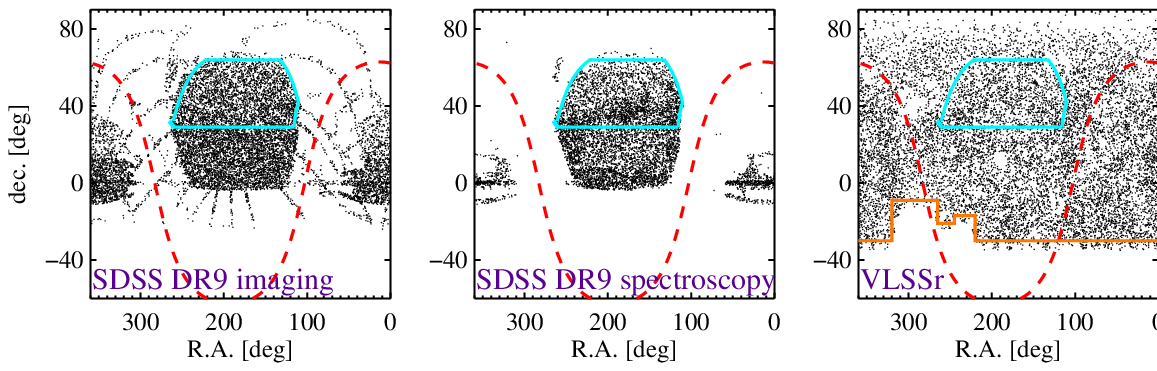}
\caption{Sky coverage of contributing surveys: SDSS DR9 imaging (photometric) survey, SDSS DR9 spectroscopic survey, and the VLSSr, as indicated by sparse sampling. Sky coverage of the other surveys is the same as in the KI08 version of the catalog, and is shown in Figure~1 of that paper. The Galactic plane is indicated by the red dashed lines. The 3269 deg$^2$ region where all surveys overlap is indicated by the cyan solid line. The orange solid line in the rightmost panel indicates the region covered by the VLSSr for the purposes of defining catalog parameters. This region contains 97\% of all sources in the VLSSr survey.
\label{fig:sky}}
\end{figure*}

\section{New VLSSr survey}

The VLSSr covers the majority of the sky north of $\delta=-40^\circ$ (i.e., the majority of the sky area covered by the original KI08 catalog). 
The sky coverage of DR9 and the VLSSr are shown in Figure~\ref{fig:sky}. (The sky coverage of the other surveys is as shown in Figure~1 of KI08.) We have defined a 3269 deg$^2$ overlap region of sky which was observed by all of the contributing surveys.

\begin{figure*}
\centering
\begin{minipage}[t]{0.45\linewidth}
\includegraphics[width=2.8in]{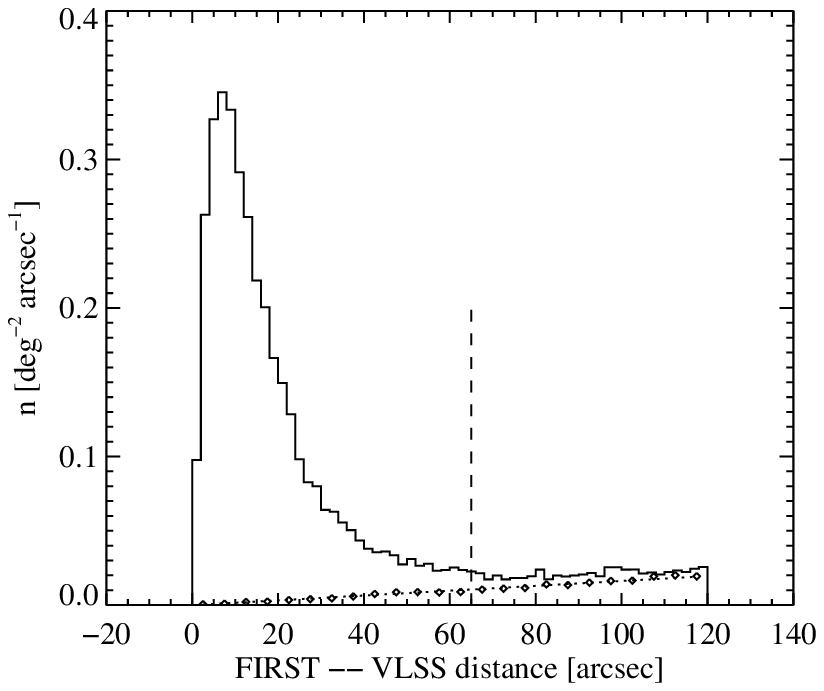}
\caption{Results of a test to determine the optimal matching radius between VLSSr positions and FIRST positions. The solid histogram indicates VLSSr counterparts within 120\arcsec\ of a FIRST source. The symbols correspond to ``random" matching, with random positions determined by off-setting the true source positions by 2 degrees in Galactic latitude.  This figure suggests that an appropriate matching radius is 65\arcsec. The VLSSr has a resolution (beam size) of 80\arcsec.
\label{fig:astrometry}}
\end{minipage}
\quad
\begin{minipage}[t]{0.5\linewidth}
\includegraphics[width=3.in]{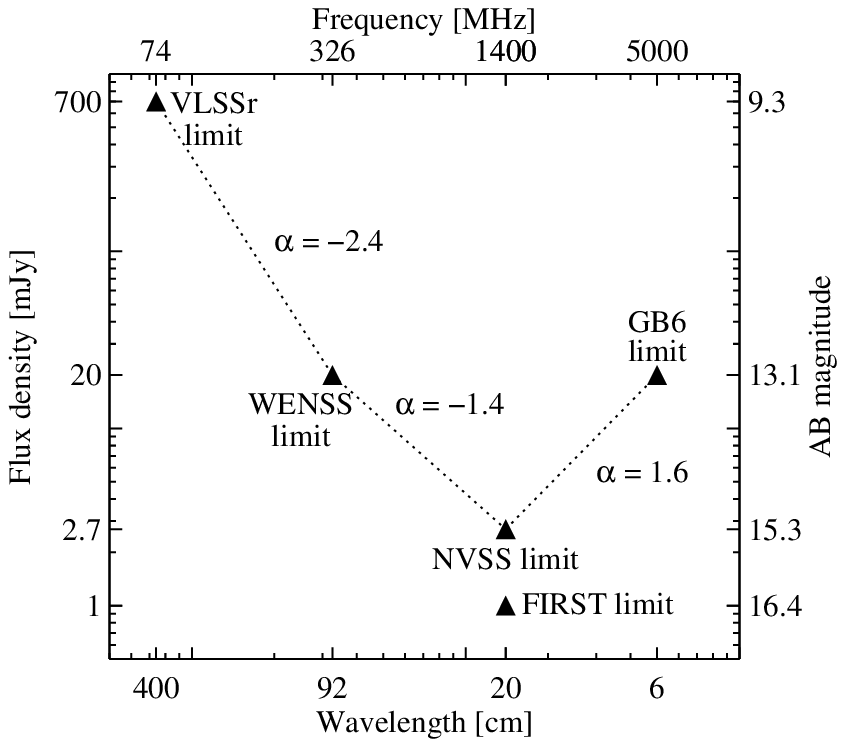}
\caption{Comparison of the flux density limits of the contributing radio surveys. The spectral indices $\alpha$ (for $f_\nu\propto\nu^\alpha$) defined at the limits of shown. Extragalactic radio sources tend to have values of $\alpha$ in the range from $-1$ (steep-spectrum) to $\sim0$ (flat-spectrum). Therefore nearly every extragalactic radio source detected in the VLSS has a counterpart in the higher-frequency surveys (in the region of overlap).}
\label{fig:surveys}
\end{minipage}
\end{figure*}

To select a matching radius between the VLSSr and the FIRST and NVSS surveys, we performed a random cross-matching analysis; the results are shown in Figure~\ref{fig:astrometry}. We cross-matched all FIRST sources within the overlap region to the VLSSr, and compared with matches to random positions, chosen by offsetting the true FIRST positions in Galactic longitude. We suggest an optimal matching radius of $\sim65^{\prime\prime}$ between VLSSr and FIRST or NVSS.

The majority of extra-galactic radio sources are flat-spectrum ($\alpha\sim0$ for $f_\nu\propto\nu^\alpha$) or steep-spectrum ($\alpha\sim<-0.5$, typically $\alpha\sim-0.8$), with increasing brightness at longer wavelengths. Therefore longer-wavelength surveys like the VLSSr are more sensitive to steep-spectrum sources while shorter-wavelength surveys are more sensitive to flat-spectrum sources. The VLSSr reaches a sensitivity of about 700 mJy at a wavelength of 4 meters, significantly brighter than the other radio surveys included here. The radio survey sensitivities are shown in Figure~\ref{fig:surveys}, with labels showing the spectral index a source would have if it were detected at the survey limits. The spectral index between the VLSSr limit and the NVSS limit ($\alpha\sim-2$) is un-physically steep, implying that most VLSSr sources should be detected in the NVSS. In fact, 98\% of VLSSr sources have an NVSS counterpart, while only about 10\% of entries in the updated catalog have a VLSSr counterpart. A selection of VLSSr sources from the updated catalog is essentially a complete, flux-limited (at 4m) sample of sources detected in all of these radio surveys, but it is a sample strongly biased toward steep-spectrum sources.

\begin{figure*}
\begin{center}
 \includegraphics[width=5in]{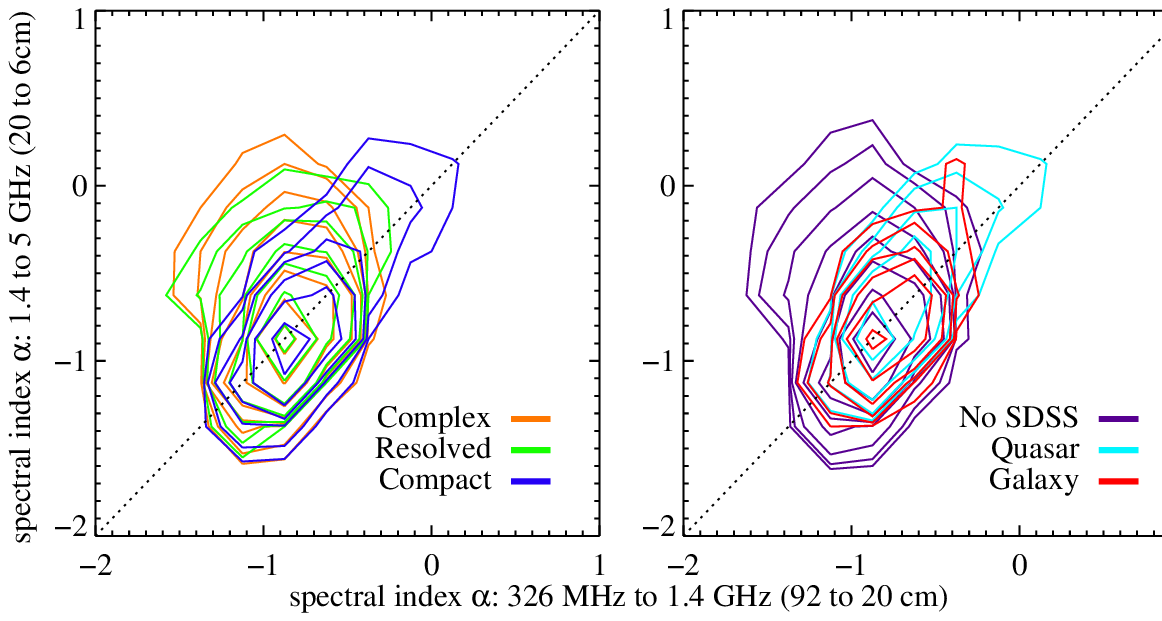}
 \caption{Spectral index distributions for the 12,000 sources detected in all five radio surveys; $\alpha<0$ implies flux density $f_\nu$ rising with increased wavelength. The dotted line indicates sources with constant spectral index. \textit{Left:} comparison by radio morphology class. Spectral indices of compact sources are typically constant from 92~cm to 6~cm, while many resolved and complex sources have spectra that flatten out toward shorter wavelengths. \textit{Right:} comparison by optical SDSS identification. Quasars tend to have flatter spectral indices than radio galaxies.}
\label{fig:contours}
\end{center}
\end{figure*}

Figure~\ref{fig:contours} illustrates morphological and spectral index characteristics of the population of sources detected in all five radio surveys (12,000 in the 3269~deg$^2$ overlap region). The left panel shows spectral index distributions of three radio morphology classes defined at 20 cm (see KI08): compact (unresolved), resolved, and complex (extended). The right panel shows spectral index distributions of three optical morphology classes: galaxy (resolved), quasar (unresolved), and optically undetected. Quasars tend to have flatter spectral indices (suggesting flat-spectrum radio-jet core sources) than galaxies (suggesting steep-spectrum radio lobes). Similarly, the compact class has more flat-spectrum sources than the resolved or complex classes. Spectral indices of compact sources are more likely to remain constant from 92~cm to 6~cm, while many resolved and complex sources have spectra that flatten out toward shorter wavelengths.

\vspace{0.5cm}

\section{Catalog access}

The updated version of the catalog (as well as the original KI08 catalog) is available at {\small \textit{http://www.atnf.csiro.au/people/Amy.Kimball/radiocat.shtml}}. The catalog parameters and format are described there, and links to online survey references are included. We have prepared a downloadable version of the complete catalog, as well as several smaller subsets of data. Subsets include the set of all sources detected by both FIRST and NVSS (580,000 entries), sources with galaxy (54,000 entries) or quasar (14,000 rows) optical spectra, and the set of isolated FIRST/NVSS sources.

\bibliography{../bibliography.bib}{}

\end{document}